\pdfoutput=1

\documentclass[aps,pra,groupedaddress,showpacs,showkeys,twocolumn,amsmath,
amsfonts,amssymb,superscriptaddress]{revtex4-2}
\usepackage{graphicx}
\usepackage[utf8x]{inputenc}
\usepackage{hyperref}
\usepackage{xcolor}
\usepackage{qcircuit}
\usepackage{physics}
\usepackage{amsbsy}

\hypersetup{
    colorlinks,
    linkcolor={red!50!black},
    citecolor={blue!60!black},
    urlcolor={blue!50!black}
}

\begin{document}

\title{Entanglement distance for arbitrary $M$-qudit hybrid systems}

\author{Denise Cocchiarella}
\affiliation{DSFTA, University of Siena, Via Roma 56, 53100 Siena, Italy}

\author{Stefano Scali}
\affiliation{Department of Physics, University of Cambridge, Cambridge CB3 0HE, United Kingdom}
\affiliation{Department of Physics and Astronomy, University of Exeter, Exeter EX4 4QL, United Kingdom}

\author{Salvatore Ribisi}
\affiliation{Centre de Physique Th\'eorique, Aix-Marseille University, Campus de Luminy, Case 907, 13288 Marseille Cedex 09, France}

\author{Bianca Nardi}
\affiliation{DSFTA, University of Siena, Via Roma 56, 53100 Siena, Italy}

\author{Ghofrane Bel-Hadj-Aissa}
\affiliation{DSFTA, University of Siena, Via Roma 56, 53100 Siena, Italy}
\affiliation{QSTAR \& CNR - Istituto Nazionale di Ottica, Largo Enrico Fermi 2,
I-50125 Firenze, Italy}

\author{Roberto Franzosi}
\email[]{roberto.franzosi@ino.it}
\affiliation{QSTAR \& CNR - Istituto Nazionale di Ottica, Largo Enrico Fermi 2,
I-50125 Firenze, Italy}


\begin{abstract}
The achievement of quantum supremacy boosted the need for a robust
medium of quantum information. In this task, higher-dimensional qudits
show remarkable noise tolerance and enhanced security for quantum key
distribution applications.
However, to exploit the advantages of such states, we need a thorough characterisation of
their entanglement. Here, we propose a measure of entanglement which can be computed either for pure and
mixed states of a $M$-qudit hybrid system. The entanglement measure is based on a
distance deriving from an adapted application of the Fubini-Study metric.
This measure is invariant under local unitary transformations and has an
explicit computable expression that we derive.
In the specific case of $M$-qubit systems, the measure assumes the physical interpretation
of an obstacle to the minimum distance between infinitesimally close states.
Finally, we quantify the robustness of entanglement of a state through
the eigenvalues analysis of the metric tensor associated with it.
\end{abstract}
\maketitle

\section{Introduction}
\label{sec:intro}
Entanglement is an essential resource for progressing in the field of
quantum-based technologies.
Quantum information has confirmed its importance in quantum cryptography
and computation, in teleportation, in the frequency standard improvement
problem and metrology based on quantum phase estimation \cite{GUHNE20091}.
The achievement of quantum supremacy \cite{arute2019quantum} together with the
rapid experimental progress on quantum control is driving the interest in entanglement theory.
Nevertheless, despite its key role, entanglement remains elusive and the problem
of its characterisation and quantification is still open
\citep{PhysRevA.95.062116,PhysRevA.67.022320}.
In time, several different approaches have been developed to quantify the variety of
states available in the quantum regime \cite{RevModPhys.81.865}.
Entropy of entanglement is uniquely accepted as measure of entanglement for pure states
of bi-partite systems \cite{PhysRevA.56.R3319}, while for the same class of mixed states,
entanglement of formation \cite{PhysRevLett.80.2245}, entanglement
distillation \cite{PhysRevA.54.3824,PhysRevLett.76.722,PhysRevLett.80.5239}
and relative entropy of entanglement \cite{PhysRevLett.78.2275} are largely acknowledged
as faithful measures. The development of quantum information theory and
the increasing experimental demand of quantum states manipulation led to
develop measures enfolding more general states.
For multi-partite systems a broad range of measures has covered pure states
\cite{PhysRevA.62.062314,briegel_PRL86_910} and mixed states \cite{PhysRevA.61.052306}
among which, a Schmidt measure \cite{PhysRevA.64.022306} and a generalisation of
concurrence \cite{PhysRevLett.93.230501} have been proposed.
In the last years, the variety of paths adopted to tackle the problem led to
estimation-oriented approaches based on the quantum Fisher information \cite{PhysRevLett.102.100401,PhysRevA.85.022321,j.aop.2019.167995}.
Due to the deep connection between the quantum Fisher information and a statistical distance
\cite{PhysRevLett.72.3439}, the geometry of entanglement has been
studied in the case of two qubits \cite{levay2004geometry}.
While the mentioned measures address mainly qubits systems, the
necessity for noise tolerance and reliability in quantum tasks opened the
way to study higher dimensional states, the qudits \cite{KHAN2006336,PhysRevA.85.032307}.
In noise-tolerant schemes, magic-state-distillation protocols outperforms their qubits 
counterparts \cite{PhysRevX.2.041021} while a proof of enhanced security for
quantum key distribution tasks is derived in \cite{PhysRevA.82.030301}.
In addition, a recent experimental realisation confirmed the superiority
of qudits in certifying entanglement in noisy environments \cite{PhysRevX.9.041042}.
At the same time, different measure of entanglement for such systems appeared, such as
a measure for highly symmetric mixed qudit states \cite{PhysRevA.95.042333} and the
$I$ concurrence in arbitrary Hilbert space dimensions \cite{PhysRevA.64.042315}.
Finally, a geometric measure for $M$-qudit pure states has been proposed in \cite{PhysRevA.80.042302}.

\begin{figure}[h]
\begin{center}
{ 
\includegraphics[width=1\linewidth]{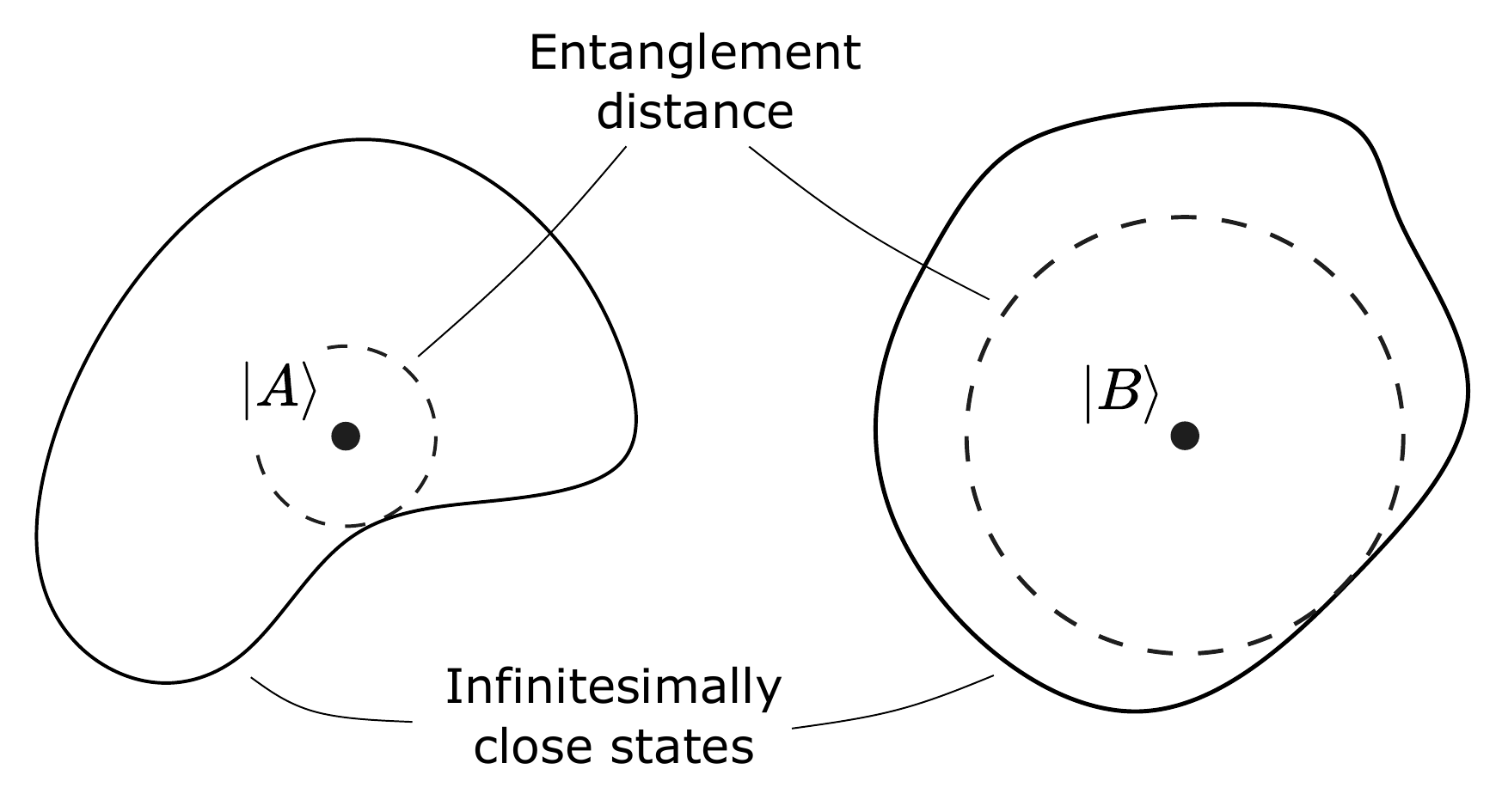}
}
\end{center}
\caption{In the specific case of 2-qubit states, the higher is the entanglement of a state the greater is its minimum distance from infinitesimally close states. In the figure, $|A\rangle$ is a low-entanglement state while $|B\rangle$ is a highly entangled state. In fact, the minimum distance (dashed line) of $|B\rangle$ from infinitesimally closed states (continuous line) is larger than the one associated with $|A\rangle$.
}
\label{draw_entanglement_distance}
\end{figure}

Following a geometric approach, in the present manuscript, we derive an entanglement
monotone \cite{Vidal_2000,PhysRevA.99.022338}, \emph{i.e.} a measure of entanglement
not increasing under local unitary transformation.
This measure can be computed either for pure and mixed
states of $M$-qudit hybrid systems.
The measure that we propose $i)$ is invariant under local unitary transformations;
$ii)$ has an explicit computable expression;
$iii)$ is derived from a tailored form of the Fubini-Study metric.
In the specific case of $M$-qubit systems, the proposed
measure $iv)$ has the structure of a distance such that the higher the entanglement
of a given state is, the greater is its minimum distance from infinitesimally close states
(see Fig. \ref{draw_entanglement_distance});
$v)$ in such case the analysis of the eigenvalues of the metric tensor associated
with the entanglement measure allows to quantify the robustness of the
entanglement of a state and determine if any states are more sensitive
to small variations than others.

\section{Entanglement distance}

A qudit, is a state in a $d$-dimensional Hilbert space ${\cal H}_d$
and a hybrid $M$-qudit is a state in the tensor product ${\cal H}:=
{\cal H}_{d_0}
\otimes {\cal H}_{d_1} \otimes \cdots  \otimes {\cal H}_{d_{M-1}}$ of
Hilbert spaces of dimension $d_0,d_1,\ldots, d_{M-1}$, respectively.
Thus, the dimension of ${\cal H}$ is $d=\prod_\mu d_\mu$.
First, we derive the entanglement measure for the case of
pure hybrid multi-qudit states, then we shall generalize this measure
to the case of mixed states.

\subsection{Pure states}

The Hilbert space 
${\cal H} = {\cal H}_{d_0}
\otimes {\cal H}_{d_1} \otimes \cdots  \otimes {\cal H}_{d_{M-1}}$
of an hybrid $M$ qudit system carries the
Fubini-Study metric~\cite{gibbons}
\begin{equation}
\langle d \psi | d \psi \rangle - \dfrac{1}{4}
|\langle \psi | d \psi \rangle - \langle d \psi | \psi \rangle|^2 \, ,
\label{F-S-metric}
\end{equation}
where $|\psi \rangle$ is a generic normalised state and $|d\psi \rangle$
is an infinitesimal variation of such state.
{The present study is aimed to endow the Hilbert space with
a Fubini Study-like metric that has the desirable property
of making it an attractive definition for entanglement
measure.}
For this reason, such distance should not be affected by
local operations on single qudits \cite{PhysRevA.57.1619,PhysRevLett.95.090503}.
As a matter of fact, the action of $M$ arbitrary $\text{SU}(d_\mu)$ local unitary 
operators $U_\mu$ ($\mu=0,\ldots,M-1$) on a given state $|s\rangle$,
generates a class of states
\begin{equation}
|U,s\rangle = \prod^{M-1}_{\mu=0} U_\mu |s\rangle
\label{Us}
\end{equation}
that share the same degree of entanglement.
For each $\mu$, $U_\mu$ operates on the $\mu$th qudit of ${\cal H}_{d_\mu}$.
Thus we define an infinitesimal variation of state \eqref{Us} as
\begin{equation}
|dU,s\rangle = \sum^{M-1}_{\mu=0} d\tilde{U}_\mu |U,s\rangle \, ,
\label{dUs}
\end{equation}
where there is no summation on the index $\mu$ and each infinitesimal
$\text{SU}(d_\mu)$ transformation $d\tilde{U}_\mu$
operates on the $\mu$-th qudit. Such infinitesimal transformation 
can be written as
\begin{equation}
d\tilde{U}_\mu = -i 
({\bf n} \cdot {\bf T})_\mu
d \xi^\mu
\label{dU}
\end{equation}
where $({\bf n} \cdot {\bf T})_\mu := {\bf n}_\mu \cdot {\bf T}_\mu $,
${\bf n}_\mu $ is an unit vector in $\mathbb{R}^{d_\mu}$, $\xi^\mu$ are
real parameters, and where
we denote by ${ T}_{\mu a}$, $a=1, \ldots,d_\mu^2-1$, the generators
of $\mathfrak{su}(d_\mu)$ algebra (see App. \ref{app1}).
From Eq. \eqref{F-S-metric}, with this choice, we obtain the following
expression for the Fubini-Study metric $g({\bf v})$,
\begin{align}
\sum_{\mu \nu} g_{\mu \nu} ({\bf v} ) &d\xi^\mu d\xi^\nu  =
\sum_{\mu \nu} 
\left(
\langle s | ({\bf v} \cdot {\bf T})_\mu
({\bf v} \cdot {\bf T})_\nu|s\rangle +
\right. \nonumber
\\ &
\left.
- 
\langle s | ({\bf v} \cdot {\bf T})_\mu
|s\rangle
\langle s |
({\bf v} \cdot {\bf T})_\nu|s\rangle
\right) d\xi^\mu d \xi^\nu \, .
\label{gmunu}
\end{align}
In the latter equation, the real unit vectors ${\bf v}_\mu$ are
derived by a rotation of the original ones according to
\begin{equation}
{\bf v}_\nu \cdot {\bf T}_\nu =
U_\nu^{ \dagger} {\bf n}_\nu \cdot {\bf T}_\nu
U_{\nu} \, ,
\label{vecrot}
\end{equation}
where there is no summation on the index $\nu$.
Focussing on a generic state $|s\rangle$,
for each $\mu=0,\ldots, M-1$, we obtain from \eqref{gmunu}
\begin{equation}
g({\bf v}_\mu)_{\mu \mu} = \sum_{ij} v_{\mu i} v_{\mu j} A_{\mu ij} \, ,
\label{gvmu}
\end{equation}
where the elements of the matrices $A_\mu$, $\mu=0,\ldots, M-1$, are
\begin{equation}
A_{\mu ij} = \langle s | T_{\mu i} T_{\mu j} |s \rangle - 
 \langle s | T_{\mu i}  |s \rangle   \langle s | T_{\mu j} |s \rangle \, .
 \label{amuij}
\end{equation}
{The proposed entanglement measure of the state $|s\rangle$
is }
\begin{equation}
E(|s\rangle) = \sum^{M-1}_{\mu = 0} \left[ \tr (A_\mu) - 2(d_\mu -1) \right]\, .
\label{emeasure}
\end{equation}

\vskip0.4cm
$E(|s\rangle)$ is a proper measure of entanglement satisfying the following
properties \cite{PhysRevLett.78.2275}:
\begin{enumerate}

\item[$i)$] The relations \eqref{first} and \eqref{third} make
the measure \eqref{emeasure}
independent from the local operators $U_\mu$. Consequently, its numerical
value is associated to the class of states generated by local unitary
transformations and not to the specific element
chosen inside the class.

\item[$ii)$] From \eqref{first} it results
\begin{equation}
\tr (A_\mu) = \dfrac{2(d^2_\mu -1)}{d_\mu}  - 
\sum^{d^2_\mu -1}_{k=1}
 \langle s | T_{\mu k}  |s \rangle^2 \, .
 \label{trAmu}
\end{equation}
Furthermore, the absolute value for the maximum eigenvalue of the set
$\{T_{\mu k}\}_k$ 
is $\sqrt{2 (d_\mu -1)/d_\mu}$ (see App. \ref{app1}), therefore we get
\begin{equation}
\tr (A_\mu) \geq \dfrac{2(d^2_\mu -1)}{d_\mu}  -  \dfrac{2 (d_\mu -1)}{d_\mu} \, .
\end{equation}
From here,
\begin{equation}
\tr (A_\mu) - 2(d_\mu -1) \geq 0 \, ,
\end{equation}
thus,
\begin{equation}
E(|s\rangle)  \geq 0 \, .
\end{equation}

\item[$iii)$] From \eqref{trAmu} we have
\begin{equation}
E(|s\rangle) \leq \sum^{M-1}_{\mu =0 } \dfrac{2(d_\mu-1)}{d_\mu}\, .
\end{equation}

\item[$iv)$] For a maximally entangled state $|s\rangle$, 
\begin{equation}
E(|s\rangle) = \sum^{M-1}_{\mu =0 } \dfrac{2(d_\mu-1)}{d_\mu}
\label{maxes}
\end{equation}
and 
\begin{equation}
\langle s | T_{\mu k}  |s \rangle =0 
\end{equation}
for each $\mu=0,\ldots,M-1$ and $k=1,\ldots,d^2_\mu -1$.

\item[$v)$]
For a fully separable state $|s\rangle= |s_0\rangle \otimes \cdots \otimes |s_{M-1}
\rangle$ from Eqs. \eqref{second} and \eqref{trAmu} we get $E(|s\rangle) =0$.

\end{enumerate}

In summary, the entanglement measure for a general hybrid qudit state
$|s\rangle$, results
\begin{equation}
E(|s\rangle) = \sum^{M-1}_{\mu =0 } \left[
\frac{2(d_\mu-1)}{d_\mu} - \sum^{d^2_\mu -1}_{k=1}
 \langle s | T_{\mu k}  |s \rangle^2
\right] \, .
\label{genpure}
\end{equation}

\subsubsection*{Qubit states}
Remarkably, in the case of a general $M$-qubit state $|s\rangle$,
\begin{equation}
\inf_{\{{\bf v}_\mu\}_\mu} 
\tr (g({\bf v}))
\end{equation}
identifies a unit vectors $\tilde{\bf v}_\nu$ for which it results
\begin{equation}
E(|s\rangle) = \tr ( g(\tilde{\bf v})) \, ,
\end{equation}
where the $\inf$ is taken over all the possible
orientations of the unit
vectors $ {\bf v}_\mu \in \mathbb{R}^{2}$.
We name entanglement metric (EM) $\tilde{g}$
the Fubini-Study metric associated to $\tilde{\bf v}_\nu$
\begin{equation}
\tilde{g} = g(\tilde{\bf v}_\nu)
\label{EM}
\end{equation}
The off-diagonal elements of $\tilde{g}$ provide
the quantum correlations between qubits.
In addition, states differing one another for local unitary
transformations have the same form of $\tilde{g}$.
In this way, the expression of EM identifies the classes of
equivalence for $M$-qubit states. 

\subsection{Mixed states}
Now, we extend the entanglement measure \eqref{emeasure} to the case of
mixed states. In order to do so, we require the measure $E$ to satisfy the following 3 conditions \cite{
PhysRevA.54.3824,
PhysRevLett.78.2275,
PhysRevLett.78.5022,
PhysRevA.62.032307,
PhysRevA.64.022306}:

\begin{enumerate}

\item[$i)$] $E(\rho) \geq 0$ and $E(\rho) = 0$ if $\rho$ is fully separable;

\item[$ii)$] $E(\rho)$ is invariant under local unitary transformation, \emph{i.e.}
$E(U \rho U^\dagger)= E(\rho)$;

\item[$iii)$] $E$ is a convex functional of the density matrix, that is
\begin{equation}
E(\alpha \rho_1 + (1-\alpha) \rho_2) \leq \alpha E( \rho_1) + (1-\alpha)E( \rho_2) 
\, ,
\end{equation}
for each $\alpha\in [0,1]$ and mixed states $\rho_1$ and $\rho_2$.

\end{enumerate}
Given a mixed state $\rho$, consider all possible ways of
expressing $\rho$ in term of pure states in the form
\begin{equation}
\rho = \sum_j p_j | \psi_j \rangle \langle \psi_j | \,  ,
\label{rho}
\end{equation}
where $p_j$ is the probability of measuring the state $|\psi_j \rangle$.
We define
\begin{equation}
E(\rho) = min \sum_j p_j E(| \psi_j \rangle ) \, ,
\label{rhomeasure}
\end{equation}
where the minimum is taken over all the possible combinations
of the form \eqref{rho}.
The conditions $i)$ and $ii)$ above, are inherited by $E(\rho)$
since the same properties hold true for $E(|s\rangle)$.
Let us verify condition $iii)$. 
Given $\rho= \alpha \rho_1 + (1-\alpha) \rho_2$, where 
$\rho_1$ ($\rho_2$) can be expressed in the form
$\sum_j p^1_j |\psi^1_j \rangle \langle \psi^1_j  |$
($\sum_j p^2_j |\psi^2_j \rangle \langle \psi^2_j  |$) in several ways.
We have $\rho = \sum_j (\alpha p^1_j |\psi^1_j \rangle \langle \psi^1_j  | + (1-\alpha )
p^2_j |\psi^2_j \rangle \langle \psi^2_j  |)$, thus
\begin{equation}
\begin{split}
\min_{\{p^1,|\psi^1\rangle, 
p^2,|\psi^2\rangle\}} \sum_j (\alpha p^1_j E(|\psi^1_j \rangle ) + (1-\alpha )
p^2_j E(|\psi^2_j \rangle ) \leq \\
 \min_{\{p^1,|\psi^1\rangle\}}  \sum_j\alpha  p^1_j E(|\psi^1_j \rangle ) +
\min_{\{
p^2,|\psi^2\rangle\}} \sum_j(1-\alpha )  p^2_j E(|\psi^2_j \rangle )
\end{split}
\end{equation}
since the minimum of a set is always less or equal to the minimum of its subsets.
\section{Examples of application}
In order to verify the efficacy of the proposed entanglement
measure, we
have first considered two families of one-parameter multi-qubit states
depending on a real parameter.
The degree of entanglement of each state depends on this
parameter and the configuration
corresponding to maximally entangled states for each of the families considered is known.
The first family of states we consider in \ref{briegrauss}, \ref{briegrauss23} 
and \ref{briegraussM},
has been introduced by 
Briegel and Raussendorf in Ref. \cite{briegel_PRL86_910}, for this
reason  we will name the elements in this family
Briegel-Raussendorf states (BRS).
The second family of states, in \ref{greenhorzeil}, is related to
the  Greenberger-Horne-Zeilinger states \cite{ghz}, since
it contains one of these states.
We will name the elements of such family Greenberger-Horne-Zeilinger--like
states (GHZLS).
{It is worth emphasizing that in Ref. \cite{briegel_PRL86_910} it has
been shown that the maximally entangled states of these two families are
not equivalent if $M \geq 4$, whereas they are equivalent if $M\leq 3$, where $M$ is
the number of qubits considered.
This fact offers us a further test for our approach to entanglement estimation.
In fact, we have found that $i)$ the entanglement measure \eqref{emeasure} provides the
same value for the maximally entangled states of both families;
$ii)$ in the case $M\leq 3$, the entanglement metric \eqref{EM} has the same
form for the maximally entangled states of the two families,
whereas for $M \geq 4$ the EMs of the maximally entangled states
of the two families are inequivalent.}
In Sec. \ref{2par},
we have considered a family of three-qubit
states depending
on two real parameters. With a suitable choice of these parameters, the state
can be fully separable or bi-separable, whereas in the generic case it is a
genuine tripartite entangled state. We will show that the proposed
entanglement measure provides an accurate description of all these
cases.  
In Sec. \ref{hqd} we have applied the entanglement measure \eqref{emeasure}
to the case of an hybrid qudit system and in Sec. \ref{qd2} to the case of two 
qutrits.

\subsection{Briegel Raussendorf states}
\label{briegrauss}
In the case of qubit, the generators ${\bf T}_\mu$ are the Pauli matrices
$\boldsymbol{\sigma}_\mu$.
We denote with $\Pi^\mu_0=(\mathbb{I}+\sigma_{\mu 3})/2$
and $\Pi^\mu_1=(\mathbb{I}-\sigma_{\mu 3})/2$ the projector operators
onto the eigenstates of $\sigma_{\mu 3}$, $|0\rangle_\mu$ (with
eigenvalue $+1$)
and $|1\rangle_\mu$ (with eigenvalue $-1$),
respectively.
Each $M$-qubit state of the BRS class is
derived by applying to the fully separable state
\begin{equation}
|r, 0 \rangle = \bigotimes^{M-1}_{\mu=0}
\dfrac{1}{\sqrt{2}}(|0\rangle_\mu+ |1\rangle_\mu)
  \, ,
\label{r0i}
\end{equation}
the non-local unitary operator 
\begin{equation}
U_0(\phi) = \exp (-i \phi H_0) = \prod\limits^{M-1}_{\mu=1}
\left(
\mathbb{I} + \alpha \Pi^\mu_0 \Pi^{\mu+1}_1 
\right) \, ,
\label{U0e}
\end{equation}
where  $H_0 = \sum^{M-1}_{\mu=1}\Pi^\mu_0 \Pi^{\mu+1}_1$
and 
$\alpha = (e^{-i\phi} -1) \, .$
The full operator \eqref{U0e} is diagonal on the states of the standard basis
$\{
|0 \cdots  0 \rangle \, , \, \,
|0 \cdots 0 1 \rangle
, \ldots ,
|1 \cdots 1 \rangle \}
$.  In fact,
each vector of the latter basis is identified by $M$ integers
$n_0,\ldots , n_{M-1} =0,1$ as
$
\ket{\{n\}} = |n_{M-1} \ n_{M-2} \quad n_0 \rangle \, 
$
and we can enumerate such vectors according to the binary integers representation
$ |k\rangle = \ket{\{n^k\}}$, with $k = \sum^{M-1}_{\mu=0} n^k_\mu 2^{\mu} $,

where $n^k_\nu$ is the $\nu$-th digit
of the number $k$ in binary representation and $k=0,\ldots,2^M-1$.
Then, the eigenvalue $\lambda_k$ of operator \eqref{U0e}, corresponding to a given
eigenstate $|k\rangle$ of this basis, results 
\begin{equation}
\lambda_k  = \sum^{n(k)}_{j=0} \binom{n(k)}{j} \alpha^j \, ,
\end{equation}
where $n(k)$ is the number of ordered couples $01$ inside the sequence of the base
vector $|k\rangle$.
For the initial state \eqref{r0i} we consistently get 
\begin{equation}
|r, 0 \rangle_M =  2^{-M/2} \sum^{2^M-1}_{k=0} |k\rangle \, ,
\label{r0}
\end{equation}
and, under the action of $U_0(\phi)$, one
obtains
\begin{equation}
\begin{aligned}
|r, \phi \rangle_M = 2^{-M/2} \sum^{2^M-1}_{k=0} 
\sum^{n(k)}_{j=0} \binom{n(k)}{j} \alpha^j
|k\rangle
\label{state-phi}
\end{aligned}
\end{equation}

For $\phi =2 \pi k$, with $k\in \mathbb{Z}$, this state
is separable, whereas, for all the other choices of the
value $\phi$, it is entangled.
In particular, in \cite{briegel_PRL86_910} it is argued
that the values $\phi = (2k+1)  \pi$, where $k\in \mathbb{Z}$,
give maximally entangled states.

\subsubsection{Fubini-Study metric for the Briegel Raussendorf states $M=2,3$}
\label{briegrauss23}

In the case of two-qubit BRS, the trace of the Fubini-Study metric is
\begin{equation}
\tr (g) = \sum^1_{\nu=0}\left[1 - c^2\left(cv_{ \nu 1}+\left(-1\right)^{\nu+1}sv_{\nu 2}
\right)^2
 \right] \, ,
\label{g2}
\end{equation}
where $c= \cos\left({\phi}/{2}\right)$ and $s= \sin\left({\phi}/{2}\right)$. \eqref{g2} is minimised with the choice $\tilde{\bf v}_\nu=\pm (c,(-1)^{\nu+1}s,0)$.
Consistently, the EM results in
\begin{equation}
\tilde{g} =
\left(
\begin{array}{cc}
s^2 & 1\\
1 & s^2
\end{array}
\right)
\label{gtilde2}
\end{equation}
and 
\begin{equation}
E(|r, \phi \rangle_2) = 2 s^2 \, .
\label{Erphi2}
\end{equation}

In the case $M=3$ and $\phi \neq (2k+1)  \pi$, with $k\in \mathbb{Z}$, the trace of $g$,
\begin{equation}
\tr (g) =\left[3 - c^2\left(c(v_{0 1} + v_{1 1} + v_{2 1})+s(v_{2 2}-v_{0 2})\right)^2
 \right] \, ,
\label{g3}
\end{equation}
is minimised with the choices $\tilde{\bf v}_0=(c,-s,0)$, $\tilde{\bf v}_1=(1,0,0)$ and
$\tilde{\bf v}_2=(c,s,0)$.
The EM and the entanglement measure in this case result to be
\begin{equation}
\tilde{g} =s^2
\left(
\begin{array}{ccc}
1 & c & -2 s^2 c^2\\
c & 1 +c^2 & c \\
- 2 s^2 c^2& c & 1
\end{array}
\right)
\label{gtilde3}
\end{equation}
and
\begin{equation}
E(|r, \phi \rangle_3) = s^2 \left( 3+ c^2 \right) \, ,
\label{Erphi2}
\end{equation}
respectively.
By direct calculation, one can verify that in the case of the maximally
entangled BRS ($\phi = (2k+1)  \pi$, $k\in \mathbb{Z}$), the choice
${\bf v}_0=(-1,0,0)$, ${\bf v}_1 = (0,0,1)$ and ${\bf v}_2=(1,0,0)$
minimizes $\tr (g)$ and the corresponding EM is the $3\times 3$
matrix of ones.

\subsubsection{Fubini-Study metric for the Briegel Raussendorf states $M>3$}
\label{briegraussM}
For a general $M$-qubit state $|r, \phi \rangle_M$, the trace of $g$ results
\begin{equation}
\begin{aligned}
\tr (g) &=\left\{ M - \sum^{M-1}_{\nu=0} \left[
v_{\nu 3}
w_{\nu 3} +
v_{\nu +} w_{\nu -} + v_{\nu -} 
 w_{\nu +}
\right]^2
 \right\} \, ,
 \end{aligned}
\label{gM}
\end{equation}
where $v_{\nu \pm} = v_{\nu 1} \pm i v_{\nu 2}$,
$c_k = 2^{-M/2} \lambda_k $, and 
\begin{equation}
\begin{array}{l}
w_{\nu -} = \sum^{2^M-1}_{k=0} 
\delta_{n_\nu^k,0} 
c^*_{k+2^{\nu}} c_k \, , 
\\
w_{\nu +} = \sum^{2^M-1}_{k=0} 
\delta_{n_\nu^k,1} 
c^*_{k-2^{\nu}} c_k \, , \\
w_{\nu 3} = \sum^{2^M-1}_{k=0} 
(-1)^{n_\nu^k} |c_k|^2 \, .
\end{array}
\label{wM}
\end{equation}
The trace is minimised by setting  
$\tilde{v}_{\nu +} ={{w}^{ \star}_{\nu -}}/{\Vert 
{\bf w}_{\nu} \Vert}$, $\tilde{v}_{\nu -} = {{w}^{ \star}_{\nu +}}/{\Vert 
{\bf w}_{\nu} \Vert}$ and $\tilde{v}_{\nu 3} = {{w}_{\nu 3}}/{\Vert 
{\bf w}_{\nu} \Vert}$.

From the latter, we get the entanglement measure for the BRS
\begin{equation}
E(|r, \phi \rangle_M) = \left( M- 
\sum^{M-1}_{\nu=0} 
\Vert {\bf w}_\nu \Vert^2 \right) \, .
\label{ErphiM}
\end{equation}

\subsection{Greenberger-Horne-Zeilinger--like states}
\label{greenhorzeil}
Now, we consider a second class of $M$-qubit states, the GHZLS, defined according to
\begin{equation}
|GHZ,\theta \rangle_M = \cos(\theta) |0\rangle + \sin(\theta) 
e^{i \varphi}|2^M-1\rangle \, .
\label{ghz}
\end{equation}
For $\theta = k \pi/2$ and $\forall \varphi$,
where $k\in \mathbb{Z}$, these states are fully separable, whereas
$\theta = k \pi/2+\pi/4$ ($\forall \varphi$) selects the maximally entangled 
states.
In this case, the trace for the Fubini-Study metric,
\begin{equation}
\begin{aligned}
\tr (g) &=M - \cos^2(2\theta)\sum^{M-1}_{\nu=0} 
(v_{\nu 3 })^2
 \, ,
 \end{aligned}
\label{g-ghzM}
\end{equation}
is minimised by the values $v_{\nu 3} =1$. Consistently, we have
\begin{equation}
\tilde{g} = \sin^2(2\theta) J_M
\end{equation}
where $J_M$ is the $M\times M$ matrix of ones. The entanglement measure
for the GHZLS results
\begin{equation}
E(|GHZ,\theta\rangle_M) = M  \sin^2(2\theta) \, .
\label{ErGHZ}
\end{equation}

We have mentioned above that in the case $M=2,3$, the maximally-entangled
BRS $|r, 2\pi k + \pi\rangle$, where $k\in \mathbb{Z}$ and the maximally
entangled GHZLS are equivalent because differing just for local unitary
transformations. In the present approach, this equivalence is caught by
the entanglement matrices. We have shown that, in the case $M=2,3$,
the EM for the maximally entangled states belonging to these two families
are identical. Furthermore, we have verified for some cases with $M>3$,
that the EMs for the maximally entangled states of the two families are
different thus confirming the results of Ref. \cite{briegel_PRL86_910}.

\subsection{Three-qubit states depending on two parameters}
\label{2par}
The last class of qubit states we consider is
\begin{equation}
\begin{aligned}
|\varphi,\gamma, \tau \rangle_3 = &\cos(\gamma) |0\rangle
[\cos(\tau) |00\rangle +
\sin(\tau) |11\rangle ] \\
+& \sin(\gamma) |1\rangle
[\sin(\tau) |00\rangle
+ \cos(\tau) |11\rangle ] \, .
\label{3qb}
\end{aligned}
\end{equation}
These states are fully separable for $\gamma = 0,\pi/2$ and $\tau = 0,\pi/2$ whereas they are bi-separable for $\tau = \pi/4$.
In this case, the trace of the Fubini-Study metric is
\begin{equation}
\begin{aligned}
\tr (g) &=\left\lbrace  3 - \cos^2(2\gamma) \cos^2(2\tau)[(v_{0 3})^2 +
(v_{1 3})^2 ] \right.\\
&\left.
 - 
[\sin(2\gamma) \sin(2\tau) v_{2 1} + \cos(2\gamma)  v_{2 3}]^2
 \right\rbrace
 \end{aligned}
\label{g-3qb}
\end{equation}
and it is minimised by the values $\tilde{\bf v}_{\nu 3} =(0,0,1)$, $\nu=0,1$ and
\begin{equation}
\begin{aligned}
\tilde{v}_{3 1} &=\dfrac{\sin(2\gamma) \sin(2\tau)}{\sqrt{
\sin^2(2\gamma) \sin^2(2\tau)+
\cos^2(2\gamma)}} \, ,\\
\tilde{v}_{3 2} &=
0 \, , \\
\tilde{v}_{3 3} &=\dfrac{\cos(2\gamma)}{\sqrt{
\sin^2(2\gamma) \sin^2(2\tau)+
\cos^2(2\gamma)}} \, .
\end{aligned}
\end{equation}
Consistently, the entanglement measure for these states results to be
\begin{equation}
E(|\varphi,\gamma, \tau \rangle_3) = 
[ 2 \sin^2(2\tau) +3 \sin^2(2\gamma) \cos^2(2\tau) ]\, .
\label{Er3qb}
\end{equation}

\subsection{Hybrid two-qudit states depending on one parameter}
\label{hqd}
As an example of application to hybrid qudit systems, we consider the
Hilbert space ${\cal H} = {\cal H}_2\otimes {\cal H}_3$, \emph{i.e.}
the product of qubit and qutrit states. 
Let us denote the elements of a basis in such Hilbert space with
$|\alpha, j \rangle$, where $\alpha = \pm$ and $j=0,1,2$ and consider
the following family of single-parameter states
\begin{equation}
|s,\theta\rangle = \cos(\theta) |+,0\rangle + \sin(\theta)|-,2\rangle \, .
\label{hy2qd}
\end{equation}
We expect the state with a higher degree of entanglement will
correspond to $\theta =\pi/4$. Note that this is not a maximally
entangled state since the component $|1\rangle$ of the second
Hilbert space is absent.
From Eq. \eqref{amuij}, we have
\begin{equation}
A_0=\left(
\begin{array}{ccc}
1 & i \cos(2\theta) & 0 \\
 -i \cos(2\theta)  & 1& 0 \\
 0 & 0 & 1- \cos^2(2\theta) 
\end{array}
\right) \, .
\end{equation}
In the case of qutrits, the generators ${\bf T}_\mu$ can be represented
with the Gell-Mann matrices.
By direct calculation, one can verify that the only non-null
matrix elements for $A_1$ are the following
\begin{align*}
(A_1)_{11} &= \cos^2(\theta)   \, , \\
(A_1)_{22} &=\cos^2(\theta)  \, , \\
(A_1)_{33} &=  \cos^2(\theta) \sin^2(\theta)  \, , \\
(A_1)_{44} &= \sin^2(\theta)  \, , \\
(A_1)_{55} &= \sin^2(\theta) \, , \\
 (A_1)_{66} &=3 \cos^2(\theta) \sin^2(\theta)   \, , \\
(A_1)_{77} &=  1  \, , \\
(A_1)_{88} &=1 \, .
\end{align*}
Thus, from Eq. \eqref{genpure} we have
\begin{equation}
E(|s,\theta\rangle) = 2 \sin^2(2 \theta) \, .
\label{eq:em_hybrid}
\end{equation}
In (\ref{eq:em_hybrid}), $\theta=\pi/4$ provides the maximally entangled state.

In the next section, we will compare entanglement measure
$E(|s,\theta\rangle) /2$ with the von Neumann entropy 
\begin{equation}
{\cal E}(\rho(\theta)) = - \cos^2(\theta) \log_2 (\cos^2(\theta))-
\sin^2(\theta) \log_2 (\sin^2(\theta))
\end{equation}
of the density matrix 
$\rho(\theta)=|s,\theta\rangle \langle s,\theta|$ associated to the same state.

\subsection{$M$-qudit states depending on two parameters}
\label{qd2}
Let us consider an $M$-qutrit system, that has a
Hilbert space ${\cal H} = {\cal H}_3\otimes \cdots \otimes {\cal H}_3$,
that is to say,
the product of $M$ qutrit states. 
We have considered the following generalisation of the GHZLS states  to qutrits,
\begin{equation}
\begin{split}
|s,\theta,\phi \rangle_M &= \sin(\theta) \cos(\phi) |0,\ldots,0\rangle + \\
&\sin(\theta) \sin(\phi)|1,\ldots, 1\rangle + \cos(\theta) |2,\ldots, 2\rangle\, ,
\end{split}
\label{2qtrit}
\end{equation}
which is a family of $2$-parameter states. We have,
\begin{align*}
(A_\mu)_{11} &= \sin^2(\theta)   \, , \\
(A_\mu)_{22} &=\sin^2(\theta)  \, , \\
(A_\mu)_{33} &= \frac{1}{4} \sin ^2(\theta ) \left(3
+\cos (2 \theta )
-2 \sin ^2(\theta ) \cos (4 \phi )\right) \, , \\
(A_\mu)_{44} &= \sin ^2(\theta ) \sin ^2(\phi )+\cos ^2(\theta )  \, , \\
(A_\mu)_{55} &= \sin ^2(\theta ) \sin ^2(\phi )+\cos ^2(\theta ) \, , \\
(A_\mu)_{66} &=  3 \sin ^2(\theta ) \cos ^2(\theta ) \, , \\
(A_\mu)_{77} &=  \sin ^2(\theta ) \cos ^2(\phi )+\cos ^2(\theta )  \, , \\
(A_\mu)_{88} &=  \sin ^2(\theta ) \cos ^2(\phi )+\cos ^2(\theta ) \, ,
\end{align*}
for $\mu=0,\dots,M-1$. Thus, it results
\begin{align}
E(|s,\theta, \phi \rangle_M) & =\frac{M}{4}  \sin ^2(\theta ) \big(
9+ \nonumber \\
&  7 \cos (2 \theta ) -2 \sin ^2(\theta ) \cos (4 \phi )\big)\, .
\label{sthetaphi}
\end{align}
In the next section we compare the entanglement measure 
$E(|s,\theta, \phi \rangle_M)/M $ of the states \eqref{2qtrit} with
the von Neumann entropy
\begin{equation}
{\cal E}(\rho(\theta, \phi) )= - a^2 \log_2 (a^2)-
b^2 \log_2 (b^2) - c^2 \log_2 (c^2) \, ,
\label{rothetaphi}
\end{equation} 
where  $\rho(\theta, \phi) = |s,\theta, \phi \rangle_{22}\langle s,\theta, \phi |$ is the density matrix
associated with the same states in the case $M=2$. Here,  $a=\sin(\theta) \cos(\phi)$, $b=\sin(\theta) \sin(\phi)$ and $c=\cos(\theta)$.

\section{Results}

\subsection{Entanglement measure}
\begin{figure}[h]
\begin{center}
{ 
\includegraphics[width=1\linewidth]{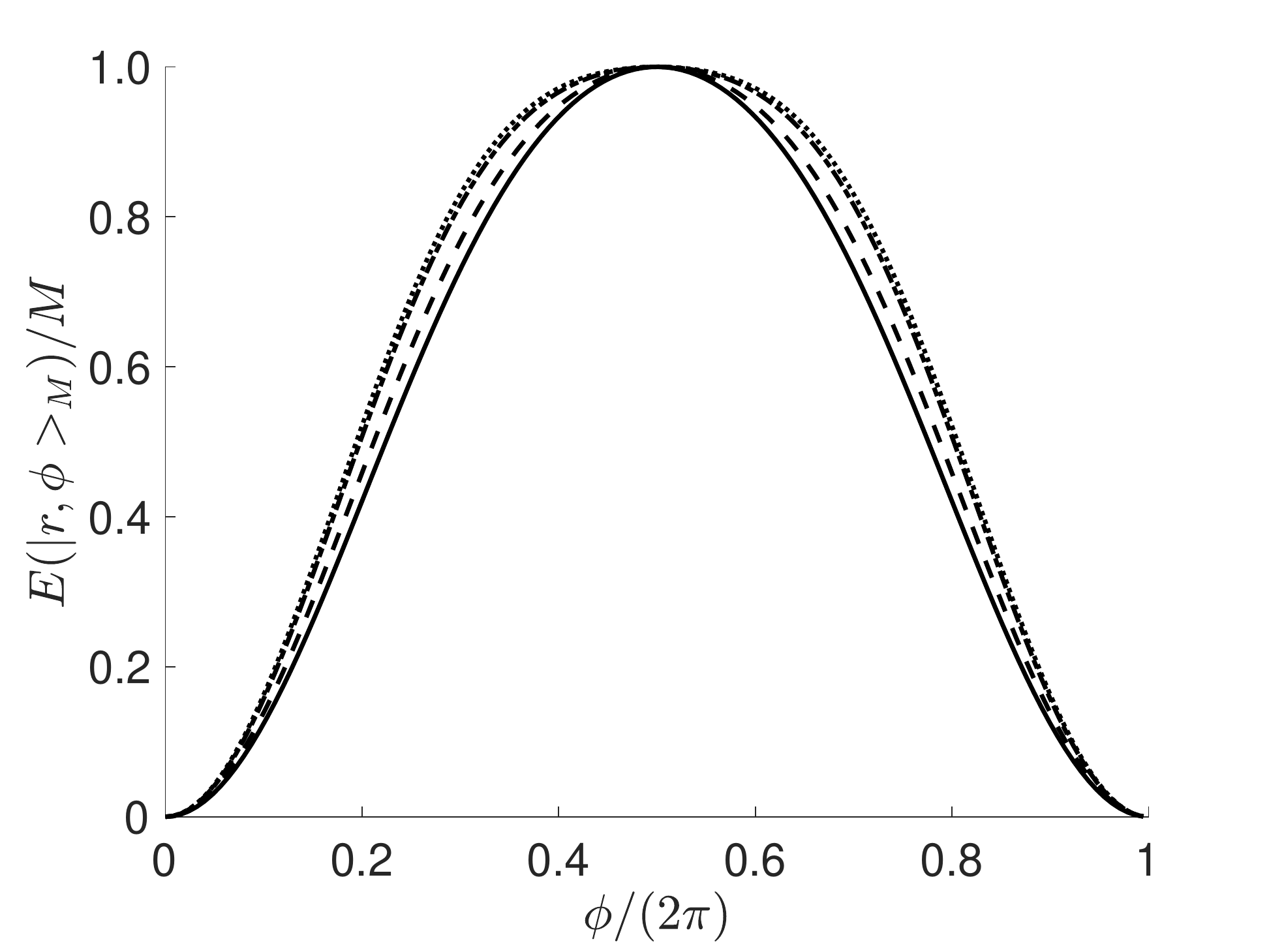}
}
\end{center}
\caption{The figure reports the entanglement measure
$E(|r, \phi \rangle_M)/M$ vs $\phi/(2\pi)$ for the states \eqref{state-phi}
in the cases $M=3$ (continuous line),
$M=4$ (dashed line), $M=7$ (dot-dashed line) and $M=9$ (dotted line).
}
\label{Fig01}
\end{figure}

In  Fig. \ref{Fig01}, we plot the measure $E(|r, \phi \rangle_M)/M$ vs $\phi/(2\pi)$
according to Eq. \eqref{ErphiM}, for the multi-qubit states
\eqref{state-phi} in the case $M=3,4,7,9$. 
Figure \ref{Fig01} shows that the proposed entanglement measure provides a
correct estimation of the degree of entanglement for the BRS in all the cases considered.
In particular, for the fully separable states ($\phi=0,2 \pi$), it is zero, whereas, for the
maximally entangled states ($\phi=\pi$), it provides the maximum possible value
for the trace, that is $E(|r, \pi \rangle_M)/M=1$. This implies that the expectation values
on the maximally entangled states of the operators
$\tilde{\bf v}_\nu \cdot \boldsymbol{\sigma}_\nu $ ($\nu=0,\ldots,M-1$) are
zero.

\begin{figure}[h]
\begin{center}
{ 
\includegraphics[width=1\linewidth]{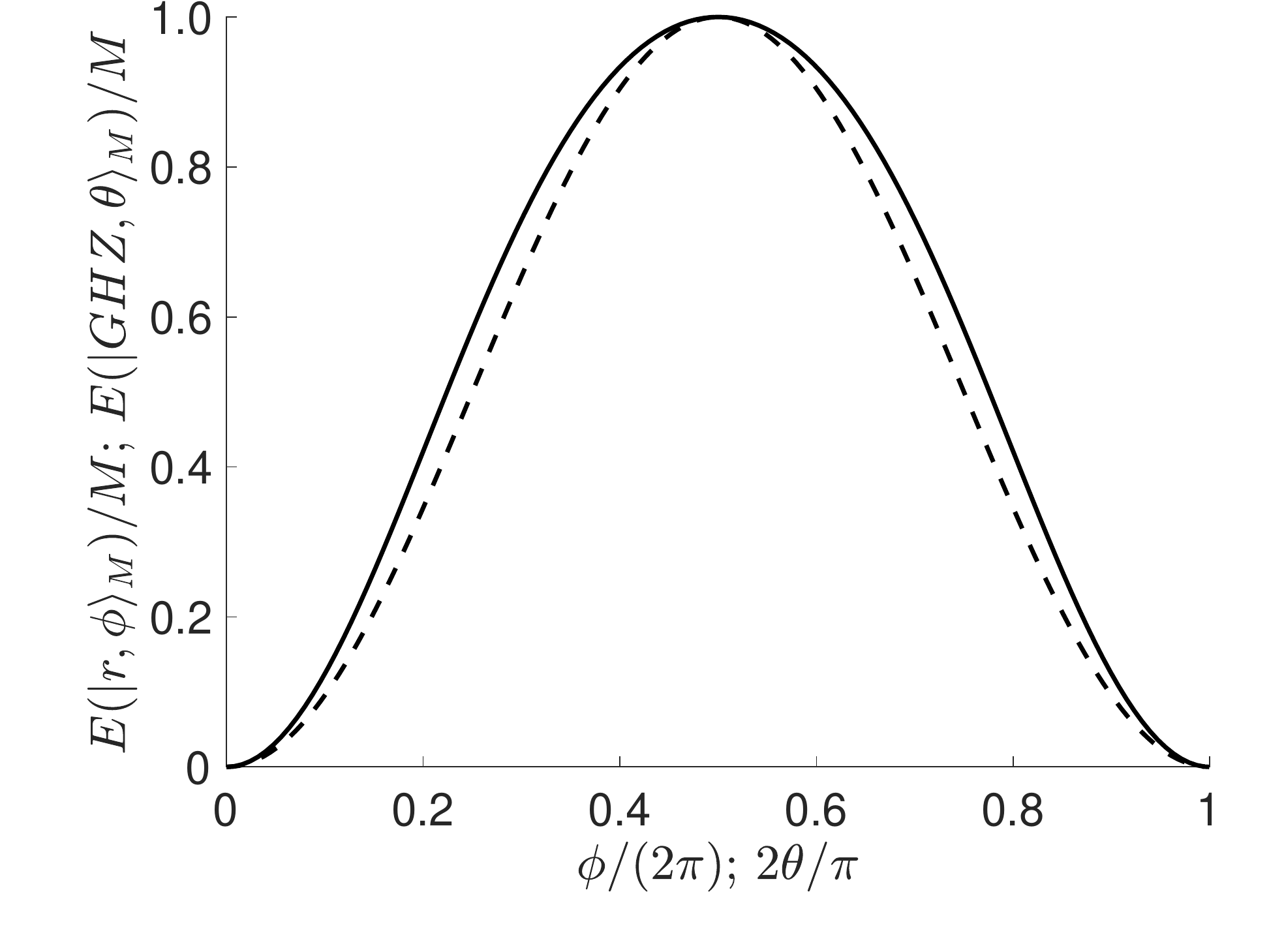}
}
\end{center}
\caption{In this figure we compare the entanglement measures
$E(|r, \phi \rangle_M)/M$ vs $\phi/(2\pi)$ for the states \eqref{state-phi} 
in continuous line, and $E(|GHZ, \theta \rangle_M)/M$ vs $2 \theta/\pi$
for the states \eqref{ghz} in dashed line,
for the case $M=3$.
}
\label{Fig02}
\end{figure}

The entanglement measure \eqref{emeasure} successfully passes also the second test
of the GHZLS for which it provides zero in the case of fully separable states
($\theta=0,\pi$), and the maximum value
 (that is $1$) in the case of the maximally entangled state ($\theta=\pi/2$). 
In figure \ref{Fig02}, we compare the curves $E(|r, \phi \rangle_M)/M$ vs $\phi/(2\pi)$ 
in continuous line and $E(|GHZ, \theta \rangle_M)/M$ vs $2 \theta/\pi$ in dashed line,
for the case $M=3$.
Even in this case, the expectation
values of the operators $\tilde{\bf v}_\nu \cdot \boldsymbol{\sigma}_\nu $
($\nu=0,\ldots,M-1$) on the maximally entangled states are zero.

\begin{figure}[h]
\begin{center}
{ 
\includegraphics[width=1\linewidth]{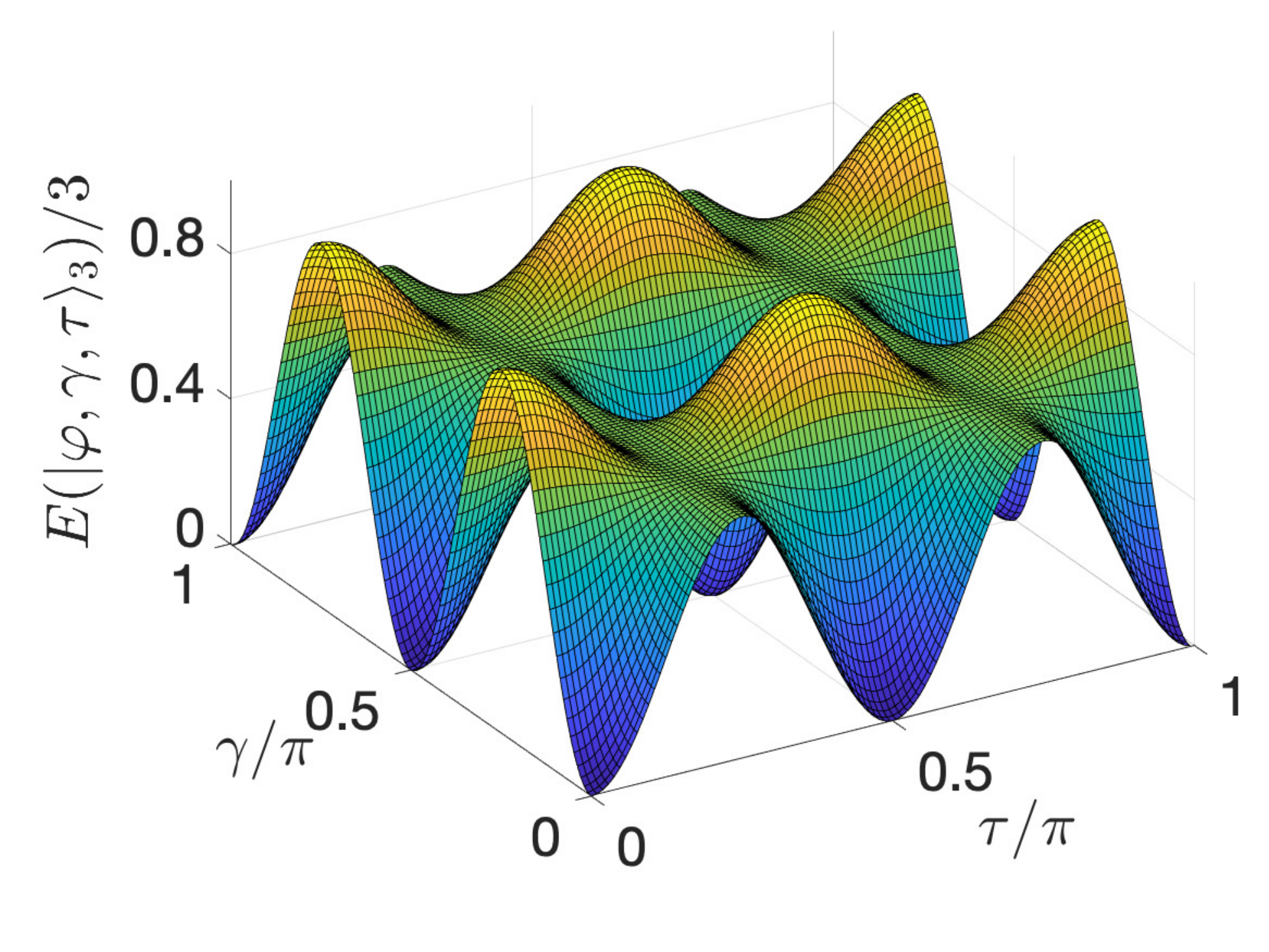}
}
\end{center}
\caption{The figure reports the three dimensional
plot of the entanglement measure $E(|\varphi,\gamma,\tau \rangle_3)/3$ 
as a function of $\gamma/\pi$ and $\tau/\pi$  for the states \eqref{3qb}.
}
\label{Fig05}
\end{figure}

In  Fig. \ref{Fig05}, we report in a 3D plot the measure
$E(|\varphi,\gamma,\tau \rangle_3)/3$ 
as a function of $\gamma/\pi$ and $\tau/\pi$ according to
Eq. \eqref{Er3qb}, for the  states \eqref{3qb}.
The measure \eqref{emeasure} catches in a surprisingly clear way the entanglement
properties of this family of states. In particular, $E(|\varphi,\gamma,\tau \rangle_3)/3$
is null in the case of fully separable states ($\gamma=0,\pi/2,\pi$ and $\tau=0,\pi/2,\pi$)
and it is maximum (with value $1$) in the case of maximally entangled states ($\gamma = \pi/4,
3\pi/4$ and $\tau= 0,\pi/2,\pi$). In addition, the case of bi-separable states ($\tau=\pi/4$) results in $0 < E(|\varphi,\gamma,\tau \rangle_3)/3 < 1$.

\begin{figure}[h]
\begin{center}
{ 
\includegraphics[width=1\linewidth]{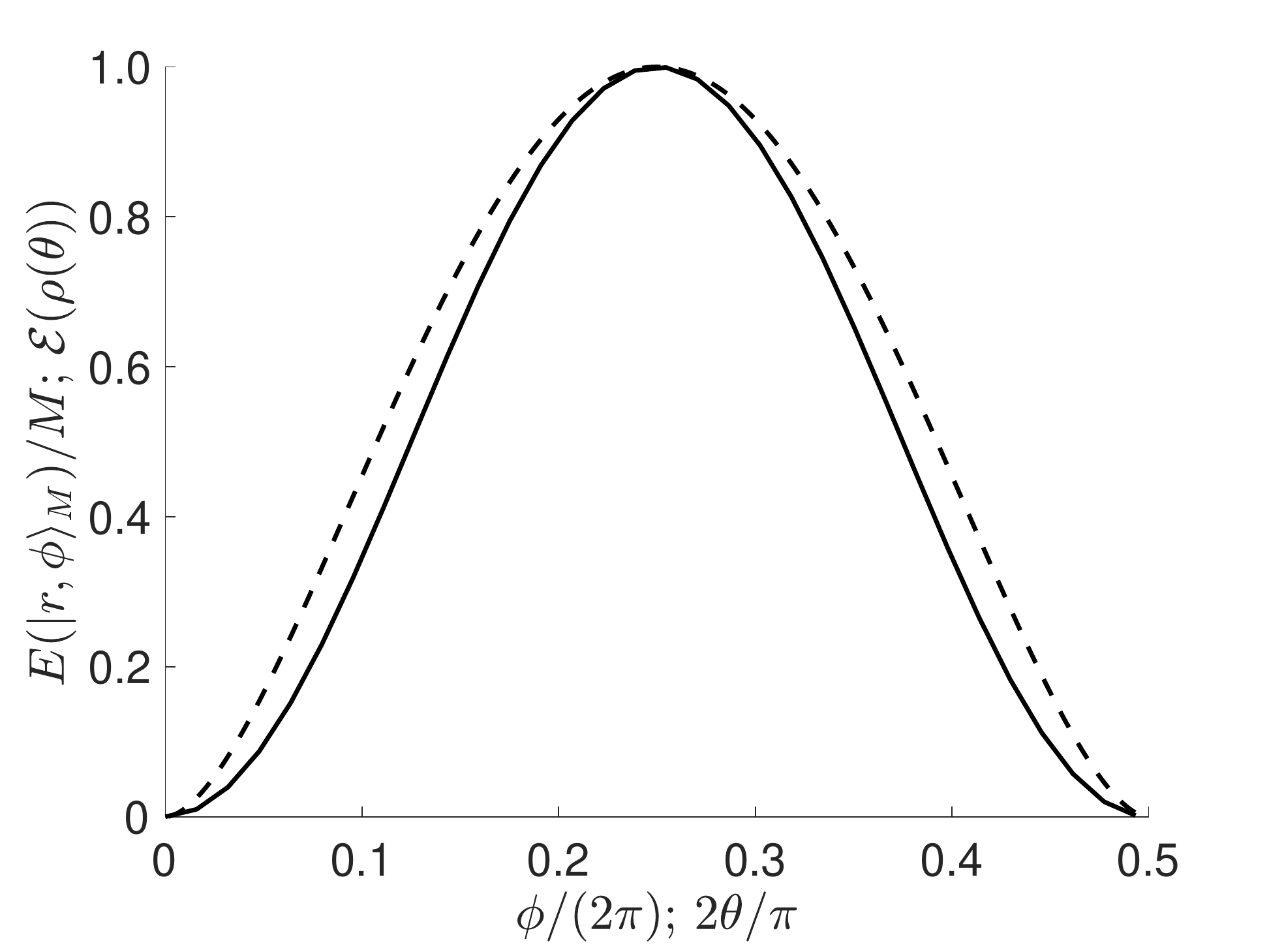}
}
\end{center}
\caption{The figure compares the entanglement measure
$E(|r, \phi \rangle_M)/M$ vs $\phi/(2\pi)$ 
in continuous line for the hybrid two-qudit states \eqref{hy2qd},
and the von Neumann entropy ${\cal  E} (\rho( \theta))$ vs $\theta/\pi$
in dashed line for the same states.
}
\label{Fig07}
\end{figure}

Figure \ref{Fig07} refers to the hybrid two-qudit states \eqref{hy2qd}. 
Here, we compare the curves of entanglement measure $E(|s, \theta \rangle)/2$
vs $\theta/\pi$ of states \eqref{hy2qd} in a continuous line, and the von
Neumann entropy  ${\cal  E} (|s, \theta \rangle)$ vs $\theta/\pi$ in
dashed line, for the same states. This figure clearly shows that, although
these two curves are different, they strongly agree in the quantification
of the entanglement of the different states. Note that the highly entangled
state associated with $\theta=\pi/4$ has
an entanglement measure of $1$, lower than the maximally entangled state
of this Hilbert space which, using \eqref{maxes}, report a value of $7/6$.

\begin{figure}[h]
\begin{center}
{ 
\includegraphics[width=1\linewidth]{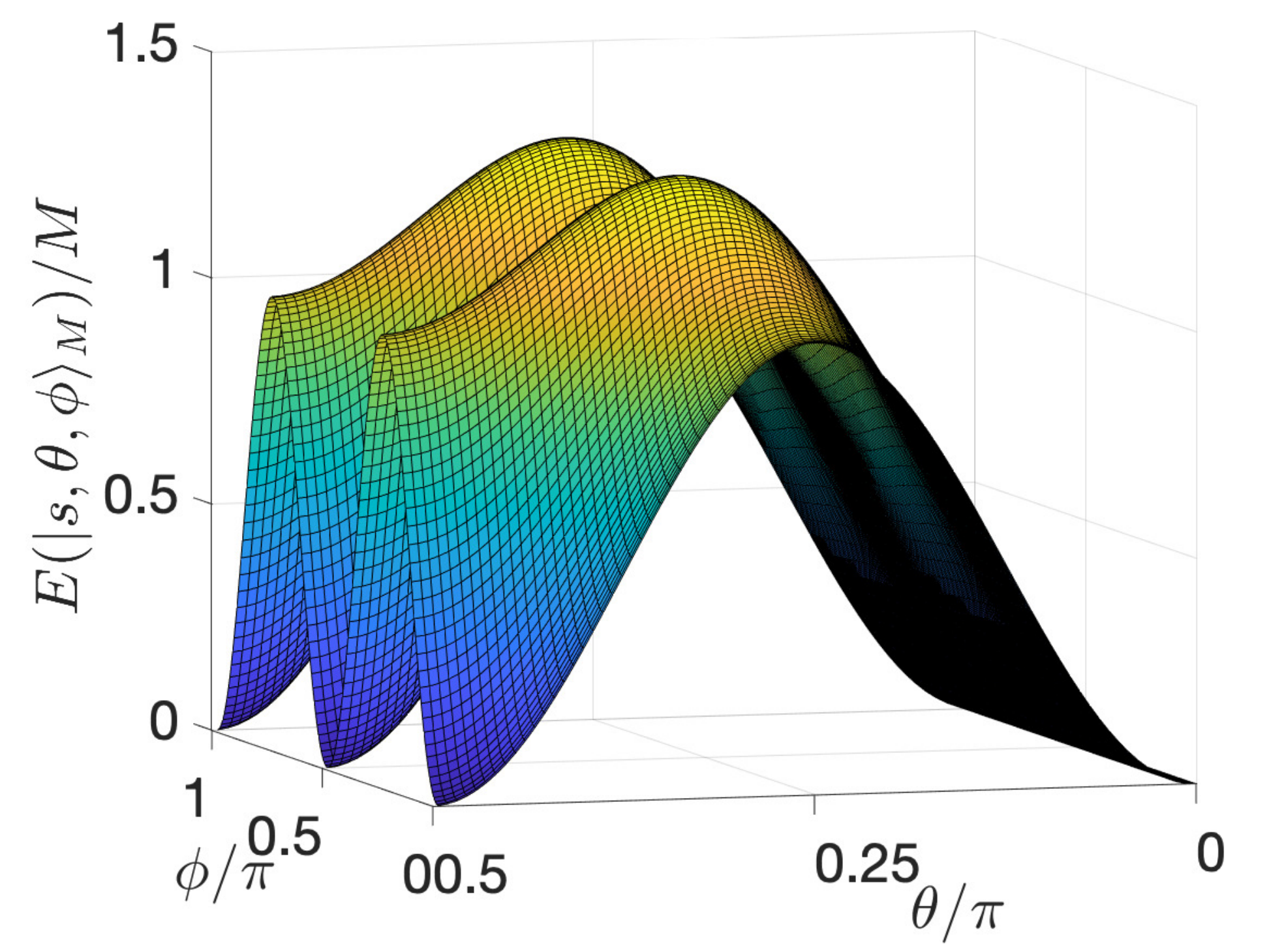}
}
\end{center}
\caption{The plot shows the entanglement measure
$E(|s,\theta,\phi \rangle_M)/M$ in \eqref{sthetaphi}
as a function of $\theta/\pi$ and $\phi/\pi$  for the states \eqref{2qtrit}.
}
\label{Fig06}
\end{figure}

In  Fig. \ref{Fig06}, we report the entanglement
measure $E(|s,\theta,\phi \rangle_M)/M$
as a function of $\theta/\pi$ and $\phi/\pi$ given in Eq. \eqref{sthetaphi},
for the multi-qubit states \eqref{2qtrit}.
Even in this example, the measure \eqref{emeasure} catches in a
surprisingly clear way the entanglement
properties of this family of multi-qudit states.
In particular, $E(|s,\theta ,\phi \rangle_M)/M$ is null in the case of
fully separable states, \emph{i.e.} for $\theta = 0$, $\forall \phi$ and
$\theta = \pi/2$, $\phi=0,\pi/2,\pi$. In case of $\phi =0,\pi$, the entanglement
measure changes over $\theta$ and shows local maximum for
$\theta=\pi/4$. For $\theta = \pi/2$, the measure changes over
$\phi$ displaying local maxima for $\phi=\pi/4, 3\pi/4$.
Furthermore, the state corresponding to
$\sin(\theta)\cos(\phi)=\sin(\theta)\sin(\phi)=\cos(\theta)=1/\sqrt{3}$
is a maximally entangled state to which corresponds an
entanglement measure \eqref{maxes} of value $4/3$.

\begin{figure}[h]
\begin{center}
{ 
\includegraphics[width=1\linewidth]{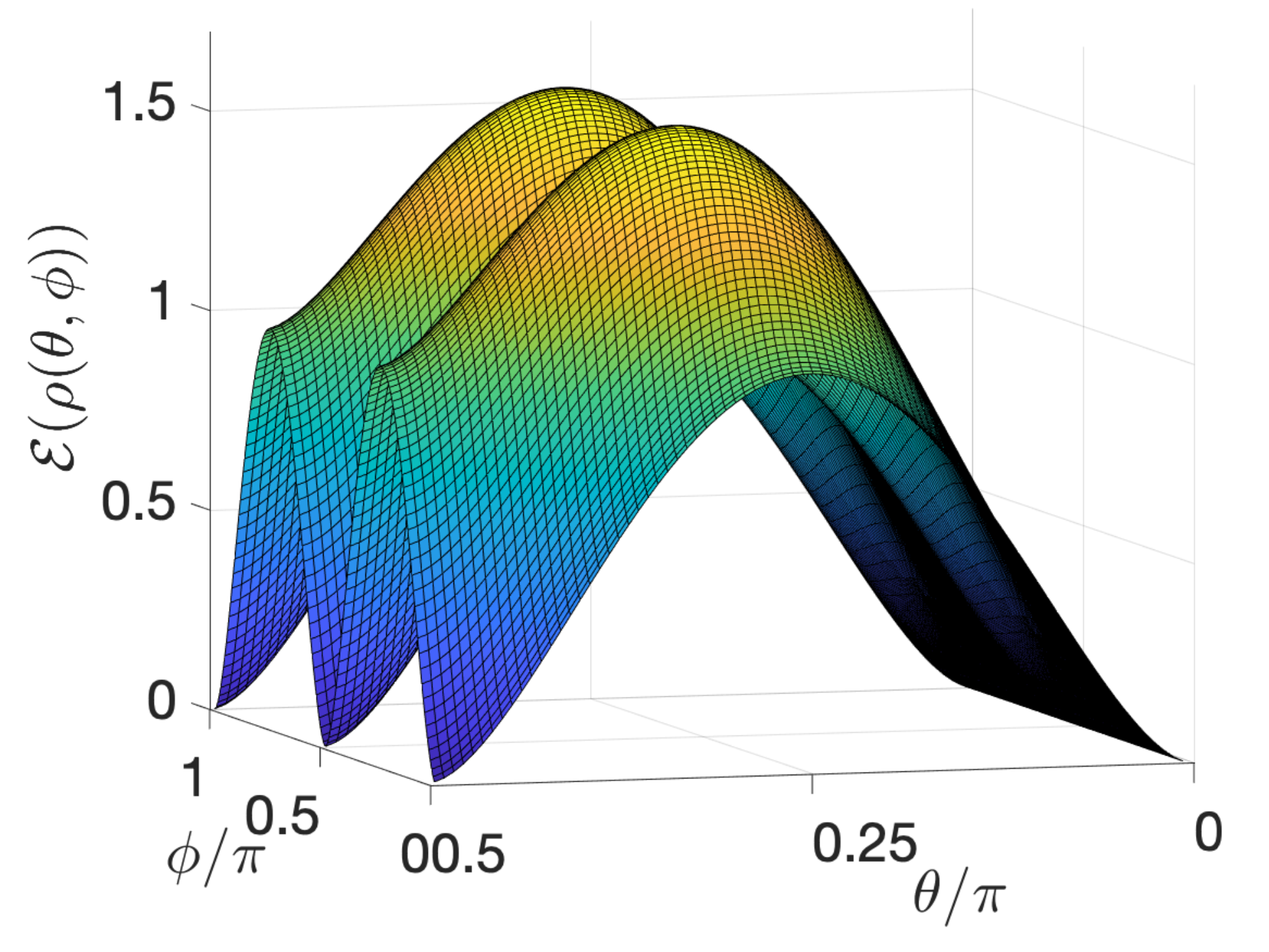}
}
\end{center}
\caption{The figure shows ${\cal E}(\rho(\theta,\phi ))$ 
as a function of $\theta/\pi$ and $\phi/\pi$  given in Eq. \eqref{rothetaphi}.
The density matrix is associated with the states \eqref{2qtrit},
$\rho(\theta,\phi) = |s,\theta,\phi\rangle_{22}\langle
s,\theta,\phi|$ in the case $M=2$.
}
\label{Fig06a}
\end{figure}

In  Fig. \ref{Fig06a}, we report the 3D plot for the von Neumann
entropy ${\cal E}(\rho(\theta,\phi ))$ (see Eq. \eqref{rothetaphi})
as a function of $\theta/\pi$ and $\phi/\pi$. The
entropy is calculated for the density matrix $\rho(\theta,\phi) = |s,\theta,\phi\rangle_{22}\langle
s,\theta,\phi|$ associated to the family of two-qudit states \eqref{2qtrit}.
The comparison between the figures \ref{Fig06} and \ref{Fig06a} clearly
shows that, although the functions $E(|s,\theta,\phi \rangle_M)/M$ and
${\cal E}(\rho(\theta,\phi ))$ are different, they fully agree, in the
entanglement estimation, for the states $|s,\theta,\phi\rangle$.

\subsection{Eigenvalues analysis for $M$-qubit states}
\begin{figure}[h]
\begin{center}
{ 
\includegraphics[width=1\linewidth]{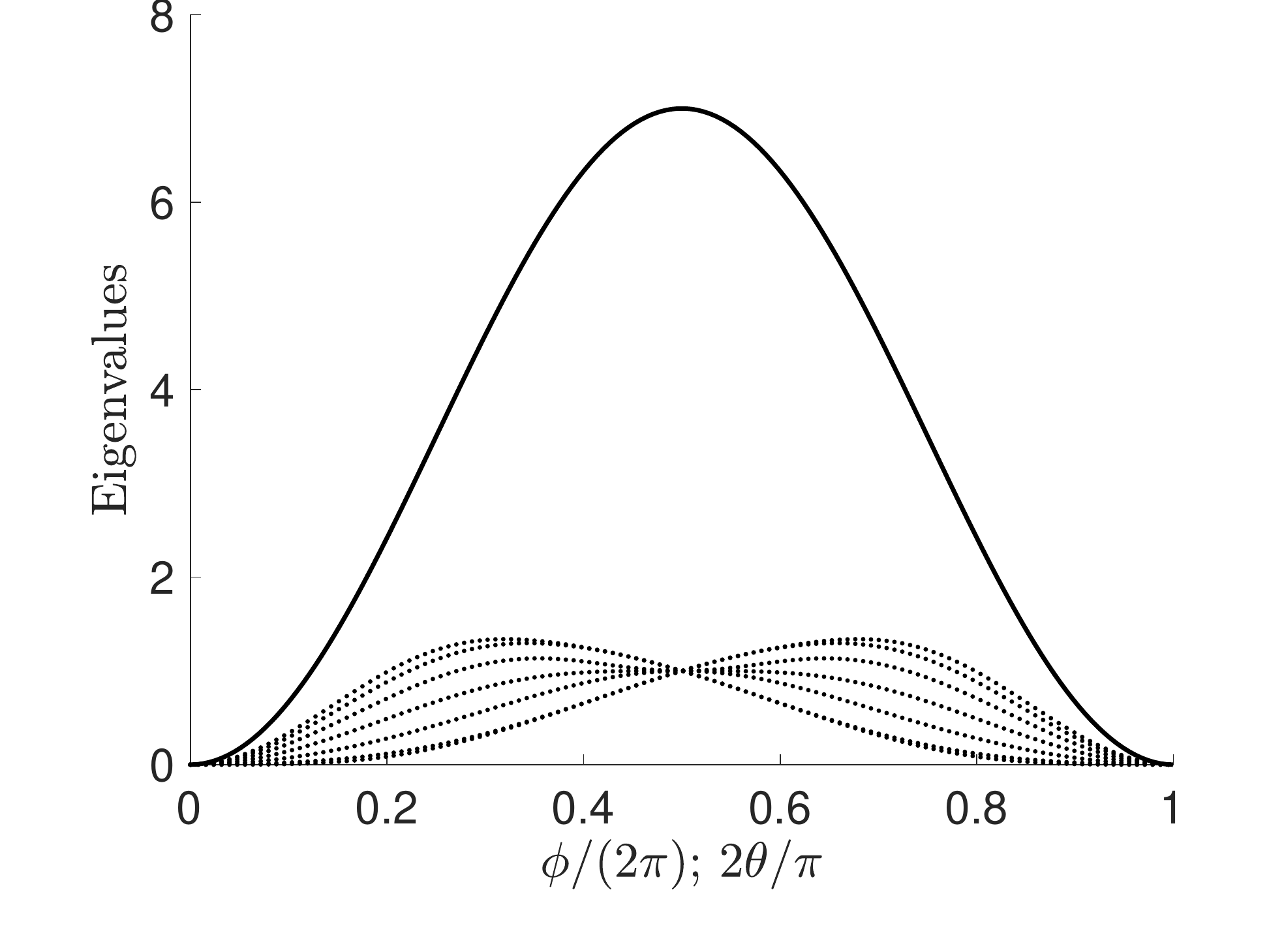}
}
\end{center}
\caption{Plot of the $\tilde{g}$ eigenvalues for the state $|r, \phi \rangle_M$ vs
$\phi/(2\pi)$ in dotted lines and the  unique not vanishing eigenvalue of 
$\tilde{g}$ for the state GHZLS vs $2 \theta/\pi$ in continuous line, for the case $M=7$.}
\label{Fig04}
\end{figure}

In the case of multi-qubit states, a further interesting characteristics
of the entanglement measure comes from
the analysis of the entanglement metric's eigenvalues.
In fig. \ref{Fig04}, we compare the plots of the eigenvalues of $\tilde{g}$ for
$|r, \phi \rangle_M$ vs $\phi/(2\pi)$ (dotted lines), with the plot of the
unique not vanishing eigenvalue of $\tilde{g}$ for GHZLS vs $2 \theta/\pi$
(continuous line), in the case $M=7$.
When $\phi\neq 0, 2\pi$ the EM of the BRS, $\tilde{g}$, has exactly
$M$ non-zero eigenvalues. On the other hand,
the GHZLS have only one non-vanishing eigenvalue. Although the
value of the latter is greater than the eigenvalues of the BRS (see Fig.
\ref{Fig04}), the GHZLS appear weak, in the sense of entanglement,
since there exist $M-1$ directions with null minimum distance between states.
This fact makes the class of the BRS robust in the sense of
entanglement. In fact, the minimum distance between states in a
random direction is greater than the minimum eigenvalue of the metric
and, therefore, greater than zero. 

\vspace{10pt}

Within the scenario that we have proposed, the entanglement has the
physical interpretation of an obstacle to the minimum distance between
infinitesimally close states. In fact, by defining the distance between a
given state represented by the vector $|U,s\rangle$ and
an infinitesimally close state associated with the vector $|d U,s\rangle$ as 
$ds^2 =\tr(g({\bf v}))dr^{2}$
where $\sum_\mu (d\xi^{\mu })^2 = dr^2$, it results
\begin{equation}
ds^2 \geq E(|s\rangle)dr^2 \, .
\end{equation}
This shows that the minimum distance density $ds^2/dr^2$, obtained by
varying the vectors ${\bf v}$, is bounded from below by the entanglement
measure $E(|s\rangle)$. 
For fully separable states, the minimum distance density is zero whereas, for
maximally entangled states, it results $M$ at the very best.
Finally, from the analysis of the eigenvalues we can investigate the sensitivity
of different states to small variations.
\begin{figure}[h]
\begin{center}
{ 
\includegraphics[width=1\linewidth]{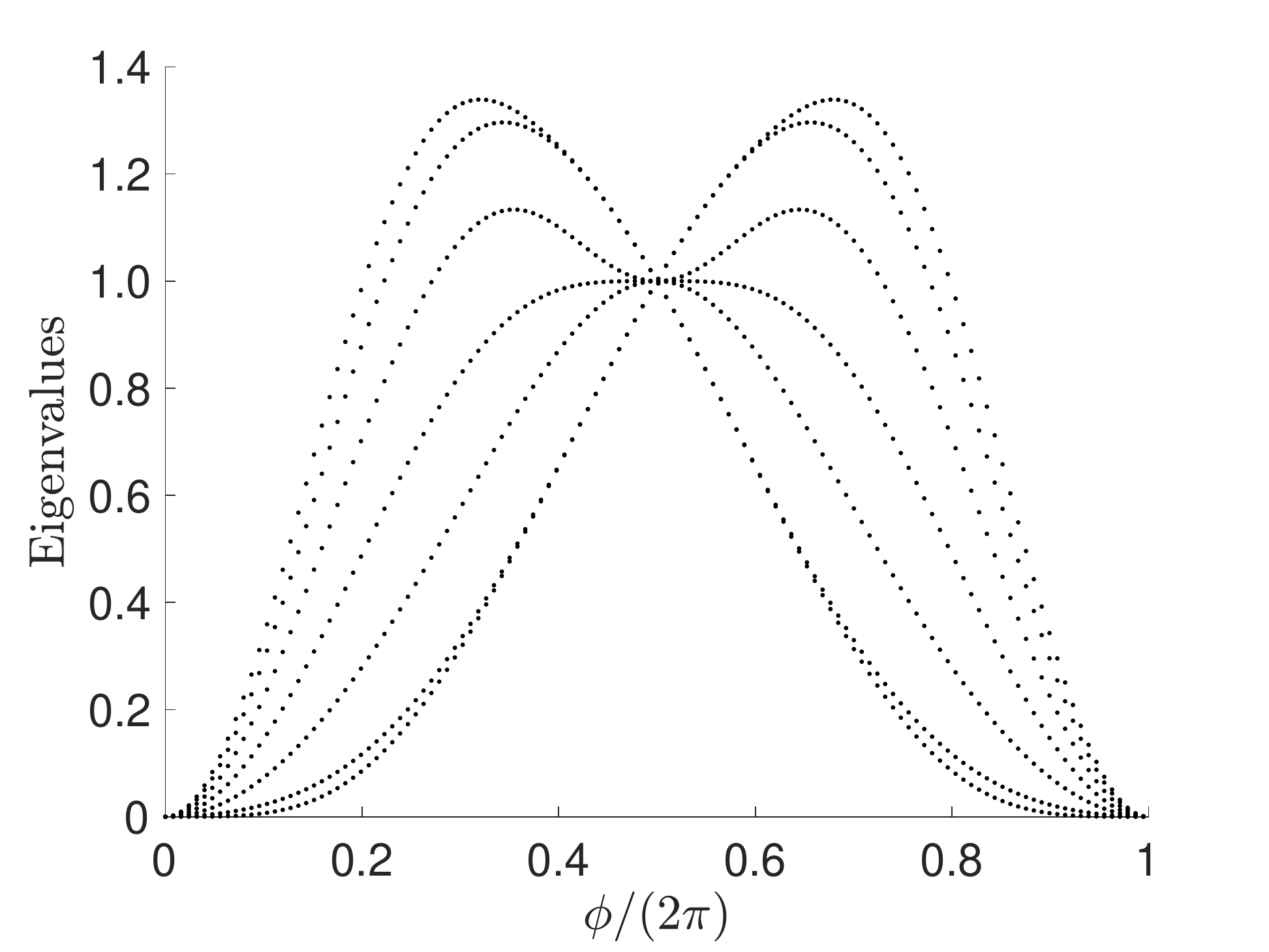}
}
\end{center}
\caption{The figure plots the $\tilde{g}$ eigenvalues for the state $|r, \phi \rangle_M$ vs
$\phi/(2\pi)$ for the case $M=7$.
}
\label{Fig03}
\end{figure}
Fig. \ref{Fig03} shows that at different points in parameter space corresponds different
state sensitivity of  $|r,\phi\rangle_7$. For instance, if we move out of $\phi=\pi/2$, 
following the eigenvector's direction corresponding to the maximum eigenvalue of $\tilde{g}$,
we find a greater distance than moving along the eigenvector's of the maximally entangled state at $\phi=\pi$.
Such analysis can be profitably used within quantum metrology applications.

\begin{acknowledgments}

R.F. thanks the support by the QuantERA project 
“Q-Clocks” and the European Commission.
S.S. is supported by the Engineering and Physical Sciences
Research Council (EPSRC) through a Doctoral Training
Grant.
\end{acknowledgments}

\appendix

\section{Generalized Gell-Mann matrices}
\label{app1}

As fundamental representation
for the generators of the algebra of $\text{SU}(d_\mu)$, we use the generalized Gell-Mann
matrices. These are the following $d_\mu^2-1$,
$d_\mu\times d_\mu$ matrices.
Let $E_{j,k}$ (for $j,k=1,\ldots, d_\mu$ ) be the matrix with $1$
as $(j,k)$-th entry and $0$ elsewhere. We define
\begin{equation}
T_{\mu \ell} = 
(E_{j,k}+E_{k,j}) \, ,
\end{equation}
 where $\ell =2(k-j)+(j-1)(2d_\mu-j)-1 $ for
$ j=1,\ldots,d_\mu-1 $, $k=j+1,\ldots,d_\mu$,
\begin{equation}
T_{\mu \ell} = -i
(E_{j,k}-E_{k,j}) \, ,
\end{equation} 
where $\ell =2(k-j)+(j-1)(2d_\mu-j) $ for
$ j=1,\ldots,d_\mu -1$, $k=j+1,\ldots,d_\mu$ and
\begin{equation}
T_{\mu \ell} = \left[\sum^k_{j=1} E_{j,j} -k  E_{k+1,k+1}
\right]\sqrt{\frac{2}{k(k+1)}} \, ,
\end{equation}
where $\ell=d_\mu (d_\mu -1)+k$ for $k=1,\ldots,d_\mu-1$.
In the case $d_\mu =2$, these generators
are given in terms of the Pauli matrices according to 
$T_{\mu 1}=\sigma_{\mu 1}$, $T_{\mu 2} = \sigma_{\mu 2}$ and
$T_{\mu 3}=\sigma_{\mu 3}$. In the case $d_\mu =3$, the 
generators are given by the standard Gell-Mann matrices.

In the general case, the following identity holds true,
\begin{equation}
\sum^{d^2_\mu -1}_{k=1} T_{\mu k} T_{\mu k} =
\dfrac{2(d^2_\mu -1)}{d_\mu} \mathbb{I}
\label{first}
\end{equation}
and, for each normalized state $| s_\mu\rangle \in {\cal H}_{d_\mu}$, it
results
\begin{equation}
\sum^{d^2_\mu -1}_{k=1} \langle s_\mu | T_{\mu k} |s_\mu\rangle^2 = 
\dfrac{2(d_\mu -1)}{d_\mu} \, .
\label{second}
\end{equation}
For each normalized state $| s\rangle \in {\cal H}$ and unitary
local operator $U_\mu:  {\cal H}_{d_\mu} \to  {\cal H}_{d_\mu}$, it
results
\begin{equation}
\begin{split}
\sum^{d^2_\mu -1}_{k=1} & \langle s |U^{ \dagger}_\mu T_{\mu_k} U_\mu 
|s\rangle^2 = \\
\sum^{d^2_\mu -1}_{k=1} & \sum^{d^2_\mu -1}_{\alpha=1} 
(n^{k }_{\alpha})^2
\langle s | T_{\mu \alpha}
|s\rangle^2 = \\
 \sum^{d^2_\mu -1}_{\alpha=1} &
\langle s | T_{\mu \alpha}|s\rangle^2 
\sum^{d^2_\mu -1}_{k=1} (n^{k}_{\alpha})^2 = \\
 \sum^{d^2_\mu -1}_{\alpha=1} &
\langle s | T_{\mu \alpha}|s\rangle^2 
\, .
\label{third}
\end{split}
\end{equation}

\hfill
\bibliography{references}

\end{document}